\pdfoutput=1
\documentclass[10pt,letterpaper]{article}
\usepackage{lineno}
\usepackage{opex3, amsmath}

\usepackage{hyperref}
\hypersetup{colorlinks=false,linkcolor=blue,citecolor=red,urlcolor=black,pdfborder={0 0 0}}
\begin{document}  
\title{High Q SiC microresonators}

\author{Jaime Cardenas$^1$, Mian Zhang$^1$, Christopher T. Phare$^1$, Shreyas Y. Shah$^1$, Carl B. Poitras$^1$,and Michal Lipson$^{1,2}$}

\address{$^1$School of Electrical and Computer Engineering, Cornell
University, Ithaca, NY, 14853\\ $^2$Kavli Institute for Nanoscale Science at Cornell, Ithaca,
NY, 14853}

\email{ml292@cornell.edu} 

\begin{abstract*} 
We demonstrate photonic devices based on standard 3C SiC epitaxially grown on silicon. We achieve high optical confinement by taking advantage of the high stiffness of SiC and undercutting the underlying silicon substrate. We demonstrate a 20 $\mu$m radius suspended microring resonator with Q=18,000 fabricated on commercially available SiC-on-silicon substrates.
\end{abstract*}

\ocis{ 230.3990 Micro-optical devices ; 140.3945   Microcavities} 

\bibliographystyle{osajnl}



\noindent Silicon Carbide (SiC) has unique properties that make it an ideal material for on-chip photonic devices, however, achieving light confinement in high quality epitaxial SiC films is challenging since these films are usually grown on silicon, which has a higher index of refraction than SiC. SiC is one of the few materials that exhibits both a high index of refraction (n=2.6), critical for compact photonic devices, and a large band gap, 2.36 - 3.3 eV depending on the polytype \cite{schaffler2001properties}, critical for minimizing absorption losses at high optical powers. The high refractive index enables highly confined photonic structures, which opens the door to compact on-chip photonic devices. The large band gap enables low loss propagation without suffering from either two or three photon absorption at telecom wavelengths from light induced free carriers \cite {yamada2012suppression}. SiC also has a high $\chi^{(2)}$ susceptibility and a high $\chi^{(3)}$  susceptibility comparable to silicon \cite{tang1991linear, chase:95}, making it an excellent material for nonlinear applications. Its crystalline nature and high stiffness make it an attractive platform for optomechanical devices \cite {ziaei2010silicon}. Of the three common silicon carbide polytypes, 4H, 6H, and 3C, the first two are only available in bulk substrates not suitable for photonic integration, while the 3C polytype is available as an epitaxially grown layer on silicon. The 3C polytype has been demonstrated to have high optical quality \cite {vonsovici2000beta}. The challenge with these films is to optically isolate the SiC (with index $\approx$2.6) from the underlying silicon (with index $\approx$3.5).

On chip photonics structures in crystalline SiC have been mainly demonstrated using challenging processes and in unusual platforms \cite {yamada2012suppression, vonsovici2000beta, tang1991optical}. Examples of such platforms and processes include SiC-on-insulator \cite {yamada2012suppression, di1996silicon}, oxygen ion implantation \cite {vonsovici2000beta}, and on SiC on sapphire \cite {tang1991optical}.  Two-dimensional slab waveguides in crystalline SiC have been demonstrated using oxygen ion implantation and SiC on sapphire \cite {vonsovici2000beta, tang1991optical} and optical confinement in three dimensions was recently demonstrated using a photonic crystal cavity on SiC-on-insulator \cite {yamada2012suppression}. Very recently micro disk photonic structures have been demonstrated in 3C SiC \cite {lusilicon}, however the micro disk geometry precludes the ability to engineer waveguide diversion, shown to be critical in nonlinear processes \cite {turner2006tailored}.

We demonstrate photonic devices based on standard 3C SiC epitaxially grown on silicon and achieve high optical confinement by taking advantage of the high stiffness of SiC and undercutting the underlying silicon substrate suspending the SiC structure in air. The high stiffness of SiC, with Young's modulus $\approx$ 400 - 450 GPa \cite {jackson2005mechanical}, enables one to suspend the structure in air using few suspension points thereby enabling high optical quality. The substrate consists of an 860 nm layer of 3C SiC on Si from NovaSiC. We deposit 300nm of Al to serve as a hard mask for etching and pattern the rings using ebeam lithography. We etch the Al hard mask in a reactive ion etcher (RIE) with a Cl$_2$ chemistry. We use a CHF$_3$ and O$_2$ chemistry in an RIE to define the SiC structure \cite {yih1997review}. Due to the chemical resistance of SiC, the etch rate was 8 nm/min. We undercut the silicon substrate using a wet KOH etch and dry the samples with a critical point dryer. Figure 1 shows an SEM of the finished device. The radius of the ring is 20 $\mu$m and the width of the waveguide is 2 $\mu$m. Two 200 nm wide spokes suspend it from the central pedestal. These spokes are designed to interact only minimally with the optical field.
\begin{figure}[htbp] \centering\includegraphics[width=104mm]{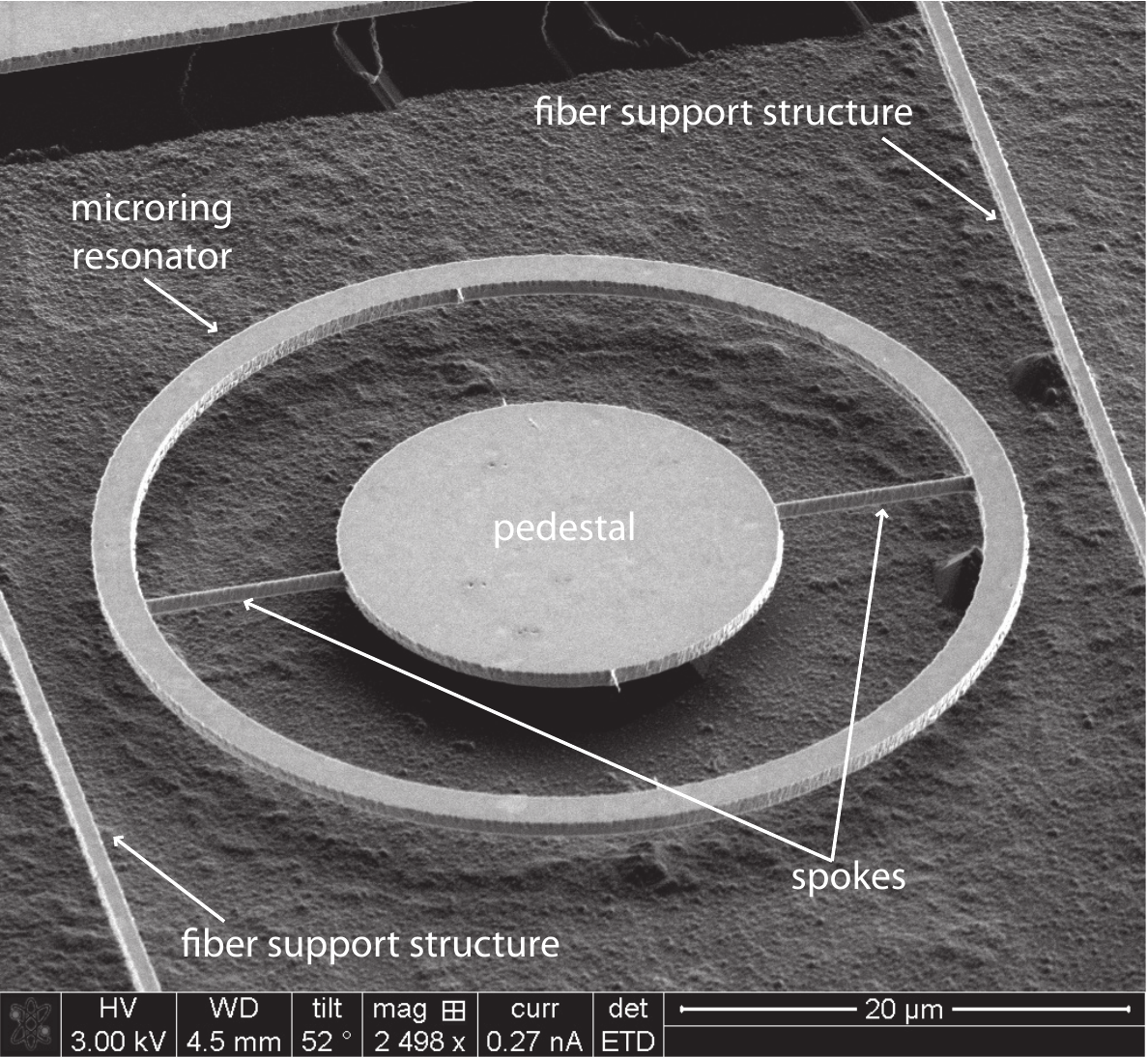}
\caption{\label{Fig1}SEM micrograph of a fabricated SiC microring resonator. SiC beams to the left and right of the resonator are support structures for the tapered fiber. The microring is 2 $\mu$m wide by 0.86 $\mu$m thick with a radius of 20 $\mu$m. The pedestal has a radius of 10 $\mu$m and spokes are 0.2 $\mu$m wide. The optical field propagates in the microring resonator.} \end{figure}

We characterize the suspended ring resonator device by measuring its optical response using a tapered fiber. We couple light from a tunable laser into the microring resonator with a tapered fiber and collect the output using a photodetector. Due to the highly confined modes of the microresonator and the large refractive index mismatch between the microresonator waveguide modes and the tapered fiber modes, the optical coupling from the tapered fiber is weak. In order to maximize coupling, we bring the tapered fiber into contact with the resonator's surface. The laser wavelength is swept while we measure the optical output. A low laser power of approximately 1 $\mu$W is used to ensure a linear response from the device. 

We measure an intrinsic quality factor of 18,000 on a 20 $\mu$m radius microring resonator.  The optical spectrum of the microresonator, as depicted in Fig. 2, has multiple resonances, corresponding to different waveguide modes resonating in the structure. Note that since some of the modes are more confined than others, some of the modes are critically coupled while other are heavily undercoupled. One can see that the mode with the highest quality factor is the one  resonating at $\lambda_0$ = 1543.2 nm (Fig. 2 inset) which is undercoupled. In order to estimate the intrinsic quality factor of this mode we fit a Lorentzian curve to the spectrum,
\begin{equation} \label{Eq1} \displaystyle T=\frac{(a-t)^2+\dfrac{4at\pi^2}{FSR^2}(\lambda-\lambda_0)^2}{(1-at)^2+\dfrac{4at\pi^2}{FSR^2}(\lambda-\lambda_0)^2},\end{equation}
where $t$ is the coupling transmission coefficient, $a$ is the round trip transmission in the cavity $a=e^{-\pi\alpha R}$, and FSR is the Free Spectral Range \cite{nitkowski2008cavity, preston2009slot}. We extract $a=0.998$ $\pm 0.0002$ and $t=0.961 \pm 0.003$, from the curve fit which results in a cavity loss $\alpha=6.2 \ {\rm cm}^{-1} \pm 0.48$. From the FSR we calculate the group index $n_g=2.77 \pm 0.04$ and using the cavity loss and group index we estimate the cavity intrinsic quality factor to be $18,000 \pm 1,300$.

\begin{figure}[htbp] \centering\includegraphics[width=131.8mm]{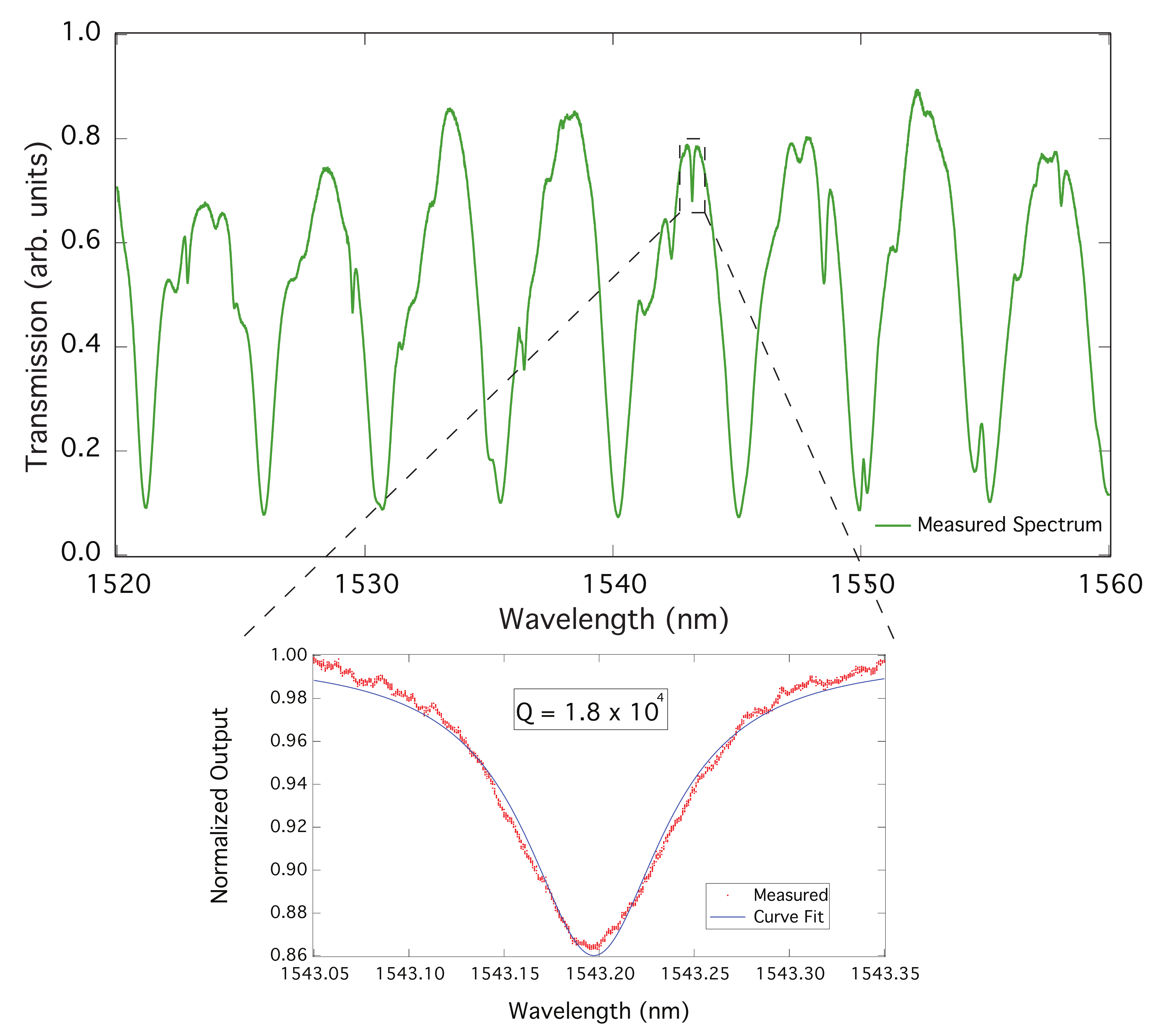}
\caption{\label{Fig2}Measured transmission spectrum for 20 $\mu$m SiC microring. The inset shows the measured resonance  and curve fit of a high quality factor resonance, \mbox{$Q = 1.8 \times 10^4$}. } \end{figure}

We identify the different modes propagating in the ring that are excited by the fiber by comparing the calculated group indices with the measured ones. In order to calculate the group indices of the different modes with high accuracy we first measure the refractive index of our 3C SiC film using spectroscopic ellipsometry. We fit a Cauchy model to the ellipsometric data and obtain an expression for the refractive index of the film: \begin{equation} \label{Eq2} n(\lambda)= 2.59-\frac{2.02\times10^{-14}}{\lambda^2} + \frac{1.90\times 10^{-26}}{\lambda^4},\end{equation} where $\lambda$ is the wavelength of light in meters. The equation is valid in the range of 1540 nm to 1560 nm. Using the film's measured refractive index, we then simulate the modes with a finite element solver. We find that the microresonator geometry supports 20 modes, however only the more delocalized ones are excited by the fiber. To extract the group indices, the modes are calculated as a function of wavelength from 1540 nm to 1560 nm. With the group index, we generate the expected resonance spectra for three of the modes. In Fig. 3, we compare the measured optical spectrum to the simulated spectra for these distinct modes. We obtain a good agreement between the measured and simulated resonances. Since the different modes present different degrees of interaction with the sidewalls and top bottom surfaces (see Fig. 3 inset), different modes present different quality factors. For example, the high Q mode, TE11, has an intrinsic Q of 18,000; while the intrinsic Q of the TE02 mode is lower at 5,000. The intrinsic Q for mode TE12 is 4,000.

\begin{figure}[htbp] \centering\includegraphics[width=131.8mm]{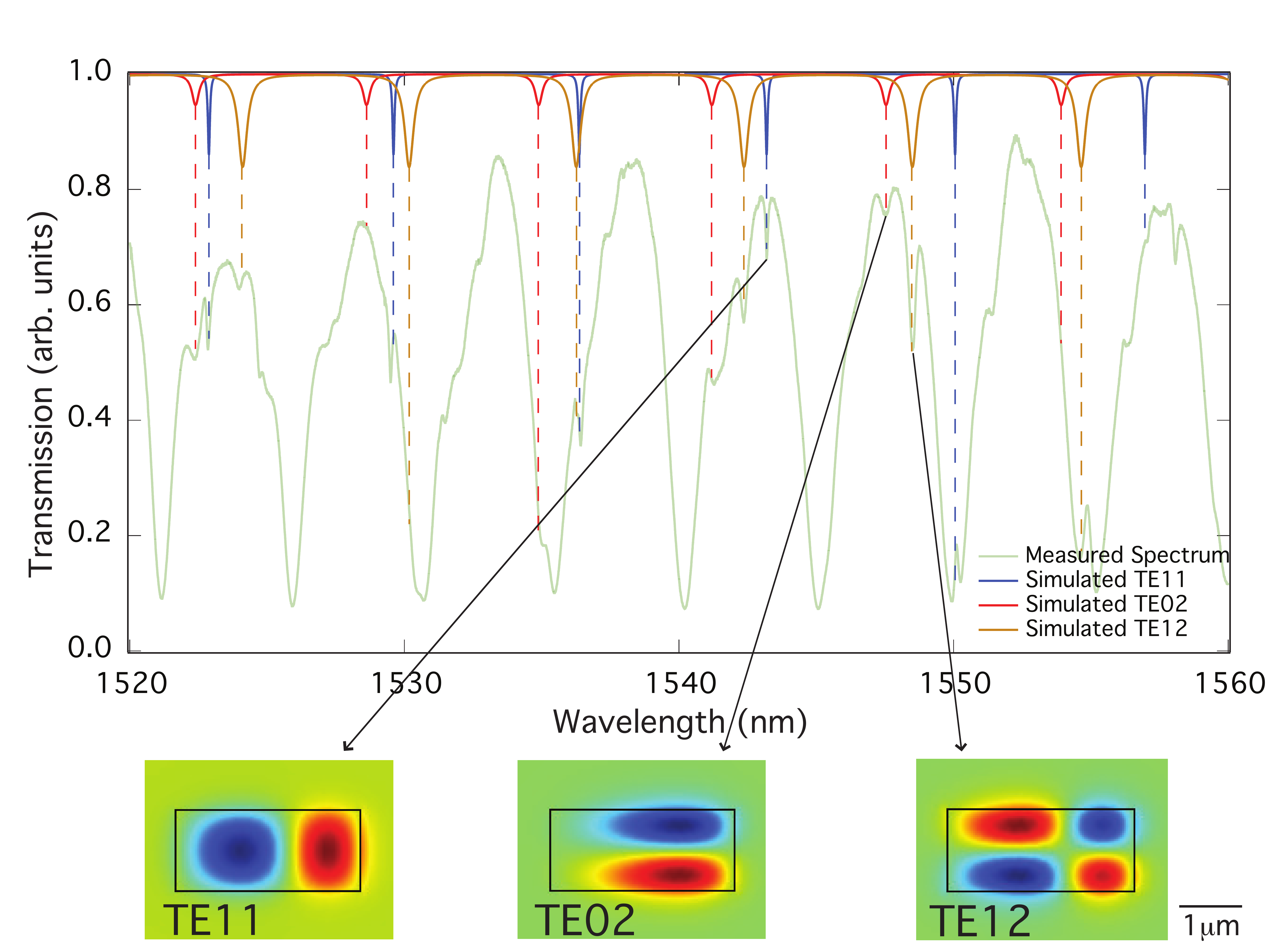}
\caption{\label{Fig3}We identify three different resonating waveguide modes in the spectrum by matching the measured spectrum with the simulated spectrum. The inset shows the major electric field component for the three quasi-TE modes identified in the spectrum: TE11, TE02, and TE12. } \end{figure}

In summary, we demonstrate a high confinement and high Q SiC ring resonator. The results show the promise of SiC as an attractive platform for on-chip photonics, optomechanics, and nonlinear devices. The suspended 3C SiC platform could enable novel optomechanical and nonlinear optical devices. Because of its high stiffness when compared to silicon, (Young's modulus of 169GPa \cite{hopcroft2010young},) we expect this approach to be scalable to long waveguides and more complex structures with high yield.

\section*{Acknowledgements}
The authors gratefully acknowledge support from the Defense Advanced Research Projects Agency (DARPA) under award \#FA8650-10-1-7064, award \#W911NF-11-1-0202 supervised by Dr. Jamil Abo-Shaeer, and from AFOSR for award \#BAA-AFOSR-2012-02 supervised by Dr. Enrique Parra. The authors wish to thank Prof. Michael Spencer for useful discussions. This work was performed in part at the Cornell NanoScale Science \& Technology Facility (a member of the National Nanofabrication Users Network) that is supported by National Science Foundation (NSF), its users, Cornell University, and Industrial Affiliates. This work made use of the Cornell Center for Materials Research Shared Facilities which are supported through the NSF MRSEC program (DMR-1120296).

\end{document}